%% file: main.tex
\documentclass[lettersize,journal]{IEEEtran}
\usepackage{amsmath,amsfonts}
\usepackage{algorithmic}
\usepackage{algorithm}
\usepackage{array}
\usepackage[caption=false,font=normalsize,labelfont=sf,textfont=sf]{subfig}
\usepackage{textcomp}
\usepackage{stfloats}
\usepackage{url}
\usepackage{verbatim}
\usepackage{graphicx}
\usepackage{cite}

\usepackage{booktabs}
\usepackage{multirow}
\usepackage{xspace}
\usepackage{subcaption}
\usepackage{tcolorbox}
\usepackage{tabularx}
\usepackage{pifont}
\usepackage{threeparttable}
\usepackage{enumitem}
\usepackage{framed}
\usepackage{xcolor}

\usepackage{breakurl}

\usepackage{hyperref}
\usepackage{tikz}

\newcommand{\tool}{\textsc{Matrix}\xspace}
\newcommand*\halfcirc[1][1ex]{%
\begin{tikzpicture}
    \draw[fill] (0,0)-- (90:#1) arc (90:270:#1) -- cycle;
    \draw (0,0) circle (#1);
\end{tikzpicture}}

\newcommand*\fullcirc[1][1ex]{\tikz\fill (0,0) circle (#1);}

\begin{document}

\title{\tool: Multi-Layer Code Watermarking via Dual-Channel Constrained Parity-Check Encoding}

\author{Yuqing Nie, Chong Wang\textsuperscript{*}, Guosheng Xu\textsuperscript{*}, Guoai Xu, Chenyu Wang, Haoyu Wang, Kailong Wang
\thanks{Yuqing Nie, Guosheng Xu, Chenyu Wang is with the School of Cyberspace Security, Beijing University of Posts and Telecommunications, Beijing, China (e-mail: jiangsha@bupt.edu.cn; guoshengxu@bupt.edu.cn; wangchenyu@bupt.edu.cn)}
\thanks{Chong Wang is with Nanyang Technological University, Singapore (e-mail: chong.wang@ntu.edu.sg)}
\thanks{Guoai Xu is with Harbin Institute of Technology, Shenzhen, China, and with Comprehensive Research Center of Electronic Information Technology in the MllT (e-mail: xga@hit.edu.cn)}
\thanks{Kailong Wang, Haoyu Wang is with the School of Cyberspace Security, Huazhong University of Science and Technology, Wuhan, China (e-mail: wangkl@hust.edu.cn; haoyuwang@hust.edu.cn)}
\thanks{\textsuperscript{*}Corresponding author.}
}


\markboth{Journal of \LaTeX\ Class Files,~Vol.~14, No.~8, August~2021}%
{Shell \MakeLowercase{\textit{et al.}}: A Sample Article Using IEEEtran.cls for IEEE Journals}


\maketitle

\begin{abstract}
Code Large Language Models (Code LLMs) have revolutionized software development but raised critical concerns regarding code provenance, copyright protection, and security. Existing code watermarking approaches suffer from two fundamental limitations: black-box methods either exhibit detectable syntactic patterns vulnerable to statistical analysis or rely on implicit neural embedding behaviors that weaken interpretability, auditability, and precise control, while white-box methods lack code-aware capabilities that may compromise functionality. Moreover, current single-layer watermarking schemes fail to address increasingly complex provenance requirements such as multi-level attribution and version tracking.

We present \tool, a novel code watermarking framework that formulates watermark encoding as solving constrained parity-check matrix equations. \tool employs dual-channel watermarking through variable naming and semantic-preserving transformations, enhancing watermark coverage across a wider range of code while ensuring mutual backup for robustness. By integrating BCH error-correction codes with solution space diversity, our approach achieves robustness against statistical analysis. Extensive evaluation on Python code generated by multiple Code LLMs demonstrates that \tool achieves an average watermark detection accuracy of 99.20\% with minimal functionality loss (0–0.14\%), improves robustness by 7.70–26.67\% against various attacks, and increases watermarking applicability by 2–6× compared with existing methods. These results establish \tool as an effective solution for complex code provenance scenarios while balancing among detectability, fidelity, and robustness.
\end{abstract}

\begin{IEEEkeywords}
Code watermarkiing, LLM watermarking, Copyright.
\end{IEEEkeywords}



\input{Chapters/intro}
\input{Chapters/background}
\input{Chapters/methodology}
\input{Chapters/evaluation}
\input{Chapters/discussion}
\input{Chapters/literature}
\input{Chapters/conclusion}

\bibliographystyle{IEEEtran}
\bibliography{paper}

\vfill

\end{document}

%% file: Chapters/intro.tex
\vspace{-0.2cm}
\section{Introduction}
Code Large Language Models (Code LLMs) have emerged as a specialized branch of LLMs, focusing on generating, understanding, and manipulating programming code. Recent advances have produced numerous powerful Code LLMs, including CodeLlama~\cite{roziere2023code}, DeepSeek-Coder~\cite{guo2024deepseek}, which have been extensively deployed in practical applications like code completion, code generation, bug fixing, and code summarization~\cite{xu2022systematic,lozhkov2024starcoder,niu2023empirical,zhou2024out,tu2023isolating,vaithilingam2022expectation}.

While Code LLMs have revolutionized software development by providing unprecedented assistance to programmers ~\cite{tan2023copilot, nam2024using, sun2024ai, wang2024teaching} and democratizing code creation for non-expert developers ~\cite{vaithilingam2022expectation, kazemitabaar2023novices}, their widespread adoption has simultaneously raised critical legal, ethical, and security concerns. A prominent issue stems from the memorization capabilities of Code LLMs, which can inadvertently reproduce copyright-protected code snippets from their training datasets without the original authors' consent~\cite{sun2022coprotector, niu2023codexleaks, al2023targeted, huang2023not}. Furthermore, due to inherent limitations in model capabilities or potential misuse by users, Code LLMs may generate defective or vulnerable code~\cite{khoury2023secure, liu2023your, mirsky2023threat, point2023opwnai, openai2023chatgptpolicy} that could be exploited for malicious purposes. The proliferation of code plagiarism presents another significant challenge, where malicious actors may appropriate open-source code, apply minor modifications, and redistribute it as original work. These challenges underscore the urgent need for robust mechanisms to accurately and effectively trace code provenance.

Similar to text watermarking~\cite{zhao2023provable, guo2024context, chang2014practical, chen2023watme}, code watermarking presents a viable approach for this purpose by embedding invisible signatures within code to identify its creators~\cite{he2023large, singh2013survey, zhu2018hidden}. However, unlike natural language, code watermarking must preserve both semantic and functional correctness. Current code watermarking schemes primarily adopt two schemes: \textbf{black-box} and \textbf{white-box} watermarking. Black-box watermarking processes the generated code to embed detectable features. 
However, these transformation-based methods typically exhibit fixed or isomorphic syntactic patterns, rendering the watermark detectable under differential analysis and making them vulnerable to theft or forgery~\cite{liu2024survey}. Moreover, end-to-end black-box watermarking methods often rely on implicitly learned embedding behaviors, making the watermarking process less interpretable and auditable, while also limiting precise control over watermark encoding and recovery.
White-box watermarking operates during the Code LLM decoding process, employing constraint strategies to select tokens suitable for adding watermark while minimizing impact on overall code quality. These methods encode watermarks by partitioning the vocabulary into green/red lists and sampling tokens exclusively from the green list. Nevertheless, these approaches lack syntactic code analysis capabilities, potentially compromising code functionality.

\begin{figure}
    \centering
    \includegraphics[width=0.9\linewidth]{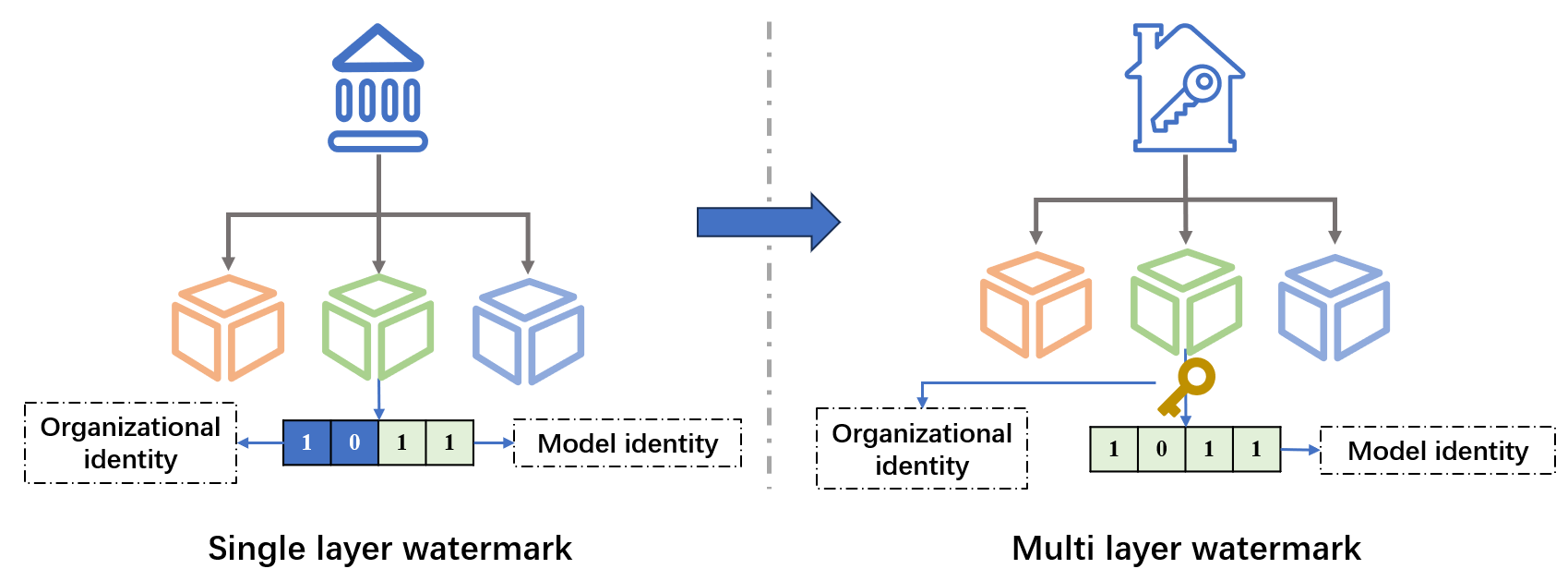}
    \caption{Single- versus Multi-layer Watermarking. Single-layer watermarking must split the sequence to encode both organization and model identities, limiting flexibility. Multi-layer watermarking enables each layer to independently use the sequence for more flexible and fine-grained attribution.}
    \vspace{-0.4cm}
    \label{fig: Multi layer watermark requirement}
\end{figure}

Furthermore, the widespread deployment of Code LLMs has introduced increasingly complex provenance requirements, including internal leak source identification, version tracking, and multi-contributor responsibility attribution~\cite{yang2023survey, li2023protecting}. Addressing these scenarios requires flexible watermarking frameworks capable of embedding customized and highly distinguishable watermarks within a unified system. However, existing methods largely adopt a coarse-grained, single-layer tracing paradigm that directly binds each attribution target to a single watermark sequence, as illustrated in Figure~\ref{fig: Multi layer watermark requirement}. Such a design is inadequate for fine-grained, multi-layer provenance tasks and fails to fully exploit the available watermark space, thereby limiting watermarking efficiency and adaptability.

\noindent\textbf{Our Work.} To address these limitations, we propose \tool, a novel framework for embedding watermark sequences into source code snippets. \tool employs a series of equivalent syntactic transformations and variable renaming operations to ensure the semantic and functional integrity of watermarked code. The framework formulates the watermark encoding process as solving constrained parity-check matrix equations to guarantee watermark detectability, where different users need only maintain their respective parity-check matrices without the costly modification of transformation rules. Furthermore, \tool utilizes a \textbf{dual-channel} mechanism: both natural channel~(through variable name substitution) and formal channel~(through semantic-preserving transformations)~\cite{yang2022natural, quiring2019misleading, allamanis2014learning, casalnuovo2020theory} to enhance watermark coverage and provide mutual backup for enhanced robustness. The framework also utilizes the error-correction space of BCH codes \cite{clark1981error, lin2021fundamentals} and the non-uniqueness of equation solutions to ensure watermark covertness against statistical analysis. Through this design, \tool achieves an optimal balance among the critical watermarking objectives, including detectability, fidelity, and robustness.

Through extensive experiments on Python code generated by multiple Code LLMs and prompt datasets, we demonstrate \tool's superior performance compared to existing baselines. \tool effectively preserves code functionality with a loss rate between 0\% and 0.14\%, achieves an average watermark detection accuracy of 99.20\%, and attains up to 100.0\% watermark information extraction accuracy. Under variable renaming attacks and LLM-based code refactoring attacks, \tool outperforms baselines with improvements of 7.70-21.72\% against various attacks, in watermark detection accuracy, respectively. Against statistical analysis, \tool exhibits more dispersed anchor states than baselines, effectively enhancing watermark indistinguishability. Moreover, \tool significantly expand the watermarking coverage by approximately 6 times compared to ACW on APPS, and twice as many on HumanEval and MBPP. These results demonstrate that \tool achieves broad applicability while effectively balancing detectability, fidelity, and robustness. Additionally, by mitigating the isomorphic syntactic patterns inherent in rule-based watermarking, \tool significantly improves robustness against statistical analysis.

\noindent\textbf{Contributions.} 
Our work's contribution is summarized below:

\begin{itemize}[leftmargin=*]
\item \textbf{Novel watermarking framework.} We formulate watermark encoding as solving constrained parity-check matrix equations, enabling flexible multi-layer provenance.
Our implementation is available at replication
package~\cite{Replication}.

\item \textbf{Fine-grained watermarking capabilities.} We introduce natural channels (variable naming) and formal channels (semantic-preserving transformations) to expand watermark coverage while providing mutual backup for enhanced robustness.

\item \textbf{BCH-enhanced covertness.} We leverage BCH error-correction codes and solution space diversity to ensure watermark covertness against statistical analysis while maintaining robustness.

\item \textbf{Superior performance.} \tool achieves average 99.20\% detection accuracy, near-perfect code functionality, 7.70-21.72\% improved robustness, and 2-6× more watermarkable samples than baselines.
\end{itemize}
\vspace{-0.2cm}

%% file: Chapters/background.tex
\section{Background and Motivation}
\subsection{LLM Code Watermarking}
With the rise of Code LLMs, code watermarking has emerged to embed creator identity. Unlike natural language, code requires semantic and functional preservation, making watermarking more challenging. The process consists of two abstract phases:

\noindent\textbf{Embedding Phase.}
Given a piece of code and an identity message $m$ (representing the code creator), we first encode it into a binary watermark sequence $w=E(m)$. The watermark is then embedded by applying a transformation function $F_{embed}$, resulting in a watermarked version:
\begin{equation}
code' = F_{embed}(code,w)
\end{equation}

\noindent\textbf{Detection Phase.} 
Given a code snippet, we attempt to extract a watermark sequence using a detection function:
\begin{equation}
\hat{w} = F_{detect}(code')
\end{equation}

Existing code watermarking methods can be broadly categorized into two types:

\noindent\textbf{Black-box Watermarking} can be modeled as a constrained transformation problem, where a target bit sequence is mapped to a set of semantics-preserving transformations applied to code. Let $C$ denote the original source code, and let $T =\{t_1,t_2,…,t_n\}$ be a set of transformations, each $t_i : C \rightarrow C $ preserving program functionality. To embed a watermark bit sequence $w \in \{0, 1\}^n$, the system applies a subset of transformations $T_w \subseteq
 T$ such that:
\begin{equation}
w_i =
\begin{cases}
1, & \text{if $t_i$ to be \textit{applied} to C,}\\
0, & \text{otherwise} 
\end{cases}
\end{equation}
\noindent The goal is to construct a transformed code $C'$ as:
\[
C' = t_{1} \circ t_{2} \circ \cdots \circ t_{k}(C), \quad \text{where } T_{i} \in T_w.
\]

In the detection phase, the system analyzes $C'$ to identify the subset of transformations $\hat{T} \subseteq T$ that have been applied, yielding an extracted bit sequence $ \hat{w} \in \{0, 1\}^n$.

\noindent\textbf{End-to-end black-box watermarking.} Unlike rule-based methods with explicit bit-to-transformation mappings, end-to-end black-box watermarking directly takes the original code C and the message m as input and generates the watermarked code $C'$:
\begin{equation}
    C' = G_{\theta}(C, m),
\end{equation}
where $G_{\theta}$ is a parameterized watermark embedding model. Correspondingly, watermark recovery is usually performed by another parameterized model $D_{\phi}$:
\begin{equation}
    \hat{m} = D_{\phi}(C').
\end{equation}

\noindent\textbf{White-box Watermarking.} Let $V$ denote the full vocabulary of the LLM,
at each token generation step $s_t$, a deterministic or pseudorandom partitioning strategy splits $V$ into two disjoint subsets: $V^{(s_t)}_{green}$ and $V^{(s_t)}_{red}$. 
Only tokens from $V^{(s_t)}_{green}$ are eligible for watermark embedding. Let $Embed(s_t)\in \{0,1\}$ denote whether the t-th position satisfies the embedding condition (e.g., the token does not affect control flow or program correctness):
\begin{equation}
Embed(s_t) =
\begin{cases}
1, & \text{if position } s_t \text{ is safe for watermarking}, \\
0, & \text{otherwise}. \\
\end{cases}
\end{equation}

\noindent This condition filters out unsafe positions where altering the output could harm functionality.
Token sampling at step $s_t$ proceeds as follows. If $Embed(s_t) = 0$, perform standard sampling. If $Embed(s_t) = 1$, constrain sampling to the green list:
\begin{equation}
\label{eq:embed}
x_t \sim \text{LM}(\cdot \mid x_{<s_t}) \big|_{V^{(s_t)}_{green}}.
\end{equation}

In the detection phase, given a generated sequence $x$, watermark detection scans positions that satisfy the embedding condition and computes the proportion of tokens $p_{\text{green}}$ that appear in the green list, 
a watermark is considered present if $p_{\text{green}}$ exceeds a predefined threshold $\tau$.

\subsection{Threat Model and Property Requirements}\label{subsec: core-requirements}

\textbf{Threat Model.} 
The code watermark is embedded into LLM-generated code before distribution, so the attacker never observes the unwatermarked version. The attacker knows the high-level framework of our scheme but not the concrete transformation rules or anchor definitions: an assumption that is practical for code watermarking~\cite{abdelnabi2021adversarial}, since full knowledge of these rules would allow an adversary to deterministically rewrite all affected constructs and trivially strip the watermark. The attacker receives only the watermarked code and aims to remove or corrupt the watermark while preserving functionality. We consider strong end-user adversaries who may apply \textit{Variable renaming}, \textit{Rule-based refactoring}, \textit{LLM-based rewriting}, or \textit{Code reformatting}.

\textbf{Property Requirements.} Based on the threat model, We outline the core requirements for a practical and resilient code watermarking system:

\noindent\textbf{Detectibility.} The watermark should be reliably embedded and accurately extracted, with minimal overhead. 
\begin{equation}
F_{detect}(F_{embed}(code, w)) = w
\end{equation}

\noindent\textbf{Robustness.} The watermark should remain intact and verifiable under the threat model. Formally:
\begin{equation}
F_{detect}(\mathcal{A}(F_{embed}(code,w))) = w
\end{equation}
where $\mathcal{A}$ denotes a transformation that preserves the functionality of the original code.

\noindent\textbf{Fidelity.} The watermarking process should preserve the program's original functionality and behavior.
\begin{equation}
semantics(F_{embed}(code, w)) = semantics(code).
\end{equation}

\noindent\textbf{Indistinguishability.} The watermark should remain indistinguishable even when an adversary observes multiple versions of the same code embedded with different messages. To prevent leakage, transformation patterns must not exhibit statistically significant correlations with the embedded bits. Otherwise, a sophisticated attacker could analyze variations across watermarked instances, such as differences in anchor activations, and correlate them with known messages to infer the encoding scheme.

\noindent\textbf{Flexibility.} The framework should support fine-grained, multi-layer attribution without requiring structural changes to the encoding system. In complex scenarios, such as identifying both an organization and an individual, the scheme should enable expressive reuse of the full watermark space, avoiding rigid bit partitioning and expensive modifications to transformation rules.

To provide a comparative overview of existing code watermarking techniques, we summarize several representative methods in Table~\ref{tab:methods}, highlighting their respective strengths and limitations in terms of the properties.

\subsection{Limitations of Existing Approaches}

\textbf{Black-box Watermarking.} While simple and modular, Black-box watermarking is mechanical and rigid: For a given message $m$, the watermarking function produces a single, fixed transformation sequence $T_w$. This one-to-one mapping,
\begin{equation}
m \xrightarrow{\text{Encode}} w \xrightarrow{\text{Transform}} T_w \xrightarrow{} C'
\end{equation}
leads to consistent syntactic patterns in $C'$. As a result, an adversary can perform statistical analysis over multiple watermarked samples to infer transformation–bit correspondences, undermining indistinguishability. Besides, each bit $w_i$ is typically tied to a single transformation $T_i$, with no alternative pathways or fallback strategies. This yields a fragile representation:
\begin{equation}
\hat{w}_i = \mathbb{I}[T_i \text{ is still present in } C']
\end{equation}
where $\hat{w}_i$ is the extracted bit. If $T_i$ is removed or altered during downstream processing or attack, $\hat{w}_i$ is corrupted. Without redundancy, the system lacks robustness against even minor perturbations.

\noindent\textbf{End-to-end black-box watermarking.} End-to-end black-box watermarking avoids explicit handcrafted rules and offers greater flexibility, but the mapping between message bits and code transformations is implicitly encoded in model parameters. As a result, such methods are weaker in interpretability, auditability, and fine-grained controllability.

\textbf{White-box Watermarking.}
As shown in Equation~\ref{eq:embed}, watermarking is only applied at positions where $Embed(s_t)=1$, indicating local suitability for watermark insertion. However, these constraints are typically based on local token-level heuristics and lack a global understanding of code structure. Consequently, even when all watermarked tokens are locally valid (i.e., $x_{s_t} \in V_{green}^{(s_t)}$ when $Embed(s_t)=1$), the resulting sequence may still violate global syntactic or semantic correctness: $x \notin C_{valid}$, 
where $C_{valid}$ denotes the set of functionally correct programs. This limitation is especially critical for code, where subtle dependencies across distant tokens can easily break compilability or functionality.

\noindent\textbf{Our Technique.} To address the limitations of existing watermarking formulations and satisfy key requirements such as fidelity, robustness, and detectability, we propose a set of design strategies. We retain semantic-preserving code transformations in black-box watermarking to ensure fidelity, while maintaining better interpretability and auditability than end-to-end embedding methods. We further introduce randomized mechanisms to improve indistinguishability by increasing variation and reducing detectability through pattern inference. Finally, we incorporate redundancy through complementary channels and solution multiplicity to improve robustness against code perturbations.

\begin{table}[]
\caption{Comparison of Existing Methods}
\label{tab:methods}
\resizebox{\columnwidth}{!}{
\begin{tabular}{c|c|c|c|c|c}
\toprule
\textbf{Type}                     & \textbf{Method}    & \textbf{Detectability} & \textbf{Robustness} & \textbf{Fidelity} & \textbf{Indistinguishability} \\ \hline
\multirow{4}{*}{Black-Box} & ToSyn~\cite{li2023protecting}     & \fullcirc       & \halfcirc           & \fullcirc        & \halfcirc            \\
                          & ACW~\cite{li2024acw}       & \fullcirc        &  \halfcirc          & \fullcirc        &  \halfcirc            \\
                          & CodeMark~\cite{li2023towards}       &  \halfcirc       &  \halfcirc          & \fullcirc        & \fullcirc             \\
                          & SrcMarker~\cite{yang2024srcmarker} &  \halfcirc        & \halfcirc          & \fullcirc        & \fullcirc            \\
                          & RoseMary~\cite{zhang2025robust}  & \fullcirc         & \fullcirc          & \halfcirc        & \fullcirc            \\ \hline
\multirow{4}{*}{White-Box} & SWEET~\cite{lee2023wrote}     & \halfcirc          & \halfcirc           & \halfcirc         & \fullcirc            \\
                          & CodeIP~\cite{guan2024codeip}    & \halfcirc         & \halfcirc           & \halfcirc         & \fullcirc            \\
                          & STA-M~\cite{mao2024watermark}    & \halfcirc          & \halfcirc           & \halfcirc         & \fullcirc            \\
                          & STONE~\cite{kim2025marking}     & \halfcirc         &  \halfcirc          & \halfcirc         & \fullcirc            \\ 
                          & MCGMark~\cite{ning2024mcgmark}     & \halfcirc         & \halfcirc           & \halfcirc         & \fullcirc            \\
\hline
\end{tabular}
}
\vspace{-0.3cm}
\end{table}

%% file: Chapters/methodology.tex
\vspace{-0.3cm}
\section{Methodology}\label{sec:METHODOLOGY}

\subsection{Definition}

We begin by defining key concepts used throughout the introduction of the methodology.

\noindent\textbf{Identity Message ($m$).} The identity message \( m \) is a unique bitstring that represents the code creator and serves as the semantic source for watermark generation. It is typically expressed as a bit sequence in \( \{0,1\}^k \), where \( k \) denotes the length of the message.

\noindent\textbf{Watermark Sequence ($w$).} The watermark sequence \( w \in \{0,1\}^l \) is a bitstring derived from the identity message \( m \) using BCH encoding. This process transforms \( m \) into a longer codeword by appending redundancy bits to enable error correction: \(w = \text{BCHEncode}(m)\).
\noindent Instead of directly using the exact BCH codeword, we randomly select \( w \) from the set of valid codewords within the error-correcting radius of \( m \). This strategy enhances indistinguishability by obscuring the one-to-one correspondence between \( m \) and \( w \). 
Specifically, a BCH code over a finite field $GF(q)$ is defined by parameters $(l, k, e)_q$, where $l$ is the watermark sequence length, $k$ is the identity message length, and $e$ is the error-correcting capability. 
The sequence \( w \) serves as the semantic watermark to be embedded and is later mapped to concrete code transformations during the embedding phase. Compared to $m$, $w$ is more robust and stealthy, due to redundancy and obfuscation introduced by the BCH encoding process.

\noindent\textbf{Transformation Rule ($t$).} A semantics-preserving transformation rule should support two inverse operations: \textit{apply} and \textit{reverse}, which transform and restore code elements, respectively. For example, we can \textit{apply} the \textit{Loop-to-Comprehension Transformation} to convert \texttt{for i in range(N): L[i] = i} into \texttt{L = [i for i in range(N)]}, and likewise \textit{reverse} the transformation to recover the original loop-based form.

\noindent\textbf{Anchor Point ($a$).}  
Given a sorted set of semantics-preserving transformation rules \( T = \{t_1, t_2, \dots, t_n\} \), an anchor point is a location in the code where a transformation operation \( t_i \) can be either \textit{applied} or \textit{reversed}.  
Each anchor point has a binary state \( s \) indicating whether it is to be \textit{transformed} with \( t_i \), defined as follows:
\begin{equation}
s =
\begin{cases}
1, & \text{if the transformation to be \textit{applied}}, \\
0, & \text{otherwise}.
\end{cases}
\end{equation}
\noindent The state \( s \) is determined by the 
transformation rule in \( T \) that can be \textit{applied} or \textit{reversed} at that location.

\noindent\textbf{State Vector ($r$).} The state vector of a given code is defined as
\begin{equation}
r = (s_1, s_2, \dots, s_p) \in \{0,1\}^p,
\end{equation}
where $p$ is the total number of anchor points in the code, and $s_i$ denotes the state of the $i$-th anchor point. This vector specifies which anchor points will be transformed using their corresponding transformation rules.

\noindent\textbf{Grouped State Vector ($c$).}  
Let a group \( G_i \) be a set of multiple anchor points. The state of \( G_i \) is set to 1 if the number of transformed anchor points in the group meets or exceeds a threshold \( \tau \); otherwise, it is set to 0. The grouped state vector is defined as
\begin{equation}
c = (c_1, c_2, \dots, c_q) \in \{0,1\}^q,
\end{equation}
where \( q \) is the total number of groups, and \( c_i \) denotes the state of the \( i \)-th group.

\vspace{-0.4cm}
\subsection{System Overview}

\begin{figure*}
    \centering
    \includegraphics[width=0.83\linewidth]{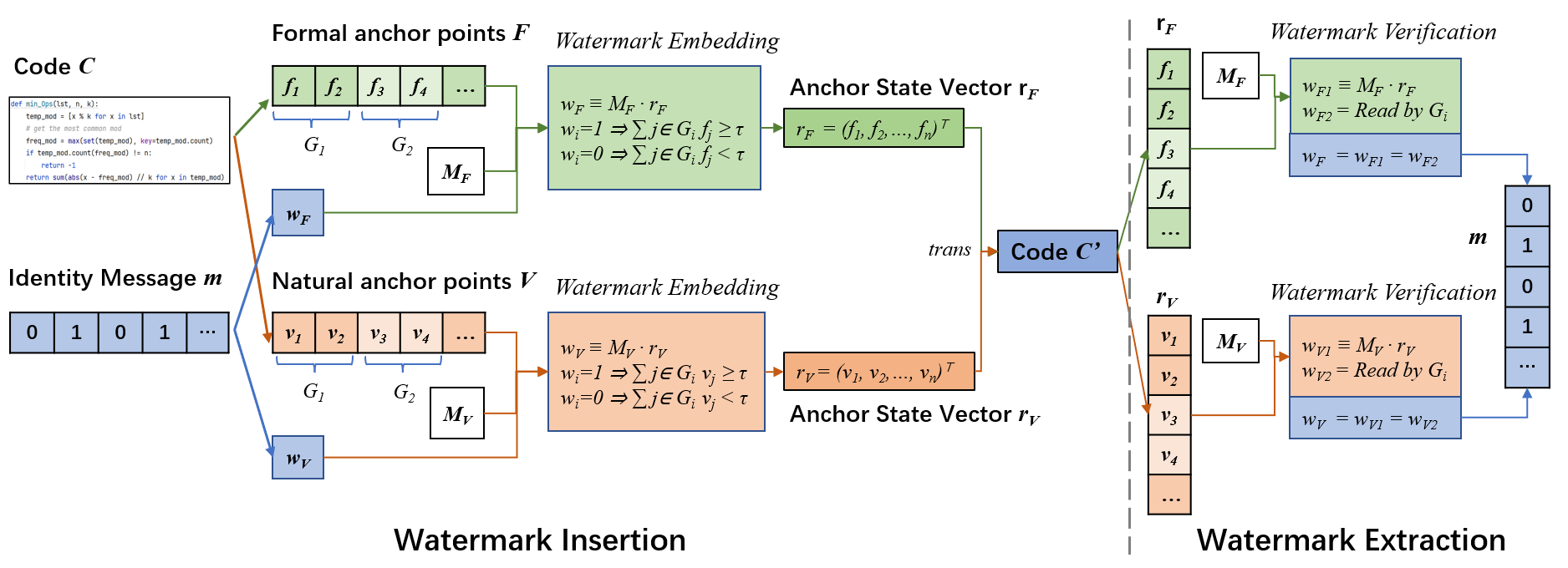}
    \caption{Overall Workflow of \tool}\vspace{-0.4cm}
    \label{fig:pipeline}
\end{figure*}

Our system, \tool, consists of a watermark insertion and a watermark extraction module, as illustrated in Figure~\ref{fig:pipeline}.
The watermarking system maintains two predefined, fixed-order transformation sets: a set of semantics-preserving transformations $T_F$ and a list of variable renaming schemes $T_V$.

During the watermark insertion process (shown in Algorithm~\ref{alg: insertion}), given a message $m$ and a code snippet $C$, appropriate watermark sequences $w$ and parity-check matrices $M \in \mathbb{F}_2^{k \times n}$ are selected for each channel based on the number of available anchor points. In both channels, the problem of computing the state vector is formulated as a Constraint Satisfaction Problem (CSP), ensuring consistency with the selected watermark sequence. The corresponding transformations are then applied according to the solution, yielding a watermarked version of the code $C'$.

In the watermark extraction phase, the input is the potentially watermarked code $C'$. \tool identifies the anchor points and their states in both channels, selects the corresponding parity-check matrices $M$, and verifies whether the extracted state vectors satisfy the parity-check equations used during insertion. If the verification succeeds, the embedded watermark sequence $w$ is recovered, and the original identity message $m$ is reconstructed.

By simply changing the configuration of the parity-check matrix, without modifying the transformation sets $T_F$ and $T_V$, \tool can flexibly support both inter-organizational and intra-organizational provenance tracing.

\begin{algorithm}[t]
\caption{Watermark Insertion}\label{alg: insertion}
\label{alg: insertion}
    \renewcommand{\algorithmicrequire}{\textbf{Input:}}  
    \renewcommand{\algorithmicensure}{\textbf{Output:}}
\small
\begin{algorithmic}[1]

\REQUIRE Original code $C$, identity message $m$, rule sets $T_F$, $T_V$, group size $\alpha$, threshold $\tau$
\ENSURE Watermarked code $C'$

\STATE \textbf{Step 1: Anchor Points Identification and Grouping}
\STATE Identify anchor sets $F$ and $V$ from $C$ using $T_F$ and $T_V$\;
\STATE Let the anchor points in $F$ and $V$ be grouped such that each group contains $\alpha$ anchor points;

\STATE \textbf{Step 2: Adaptive Embedding Capacity Estimation}
\FOR{each anchor set \{$F$, $V$\}}
    \STATE $n \leftarrow$ number of anchors
    \IF{$n \ge \alpha \cdot |w|$}
        \STATE $w \leftarrow \text{BCHEncode}(m)$
    \ELSIF{$\alpha < n < \alpha \cdot |w|$}
        \STATE $w \leftarrow m$
    \ENDIF
    
    \STATE select M that $\mathrm{rows}(M) = |w|$\;
    \IF{$n < \alpha$}
        \STATE $w \leftarrow$ organization code
        \STATE $M \leftarrow$ unit parity-check matrix
    \ENDIF
\ENDFOR

\STATE \textbf{Step 3: State Vector Computation}
\STATE Set grouped state vector $c$ such that $c_i = w_i$ for all $i$\;
\STATE Solve:
\begin{equation}
\begin{cases}
M \cdot r \equiv c \pmod{2} \\
c_i = 1 \Rightarrow \sum_{j \in G_i} s_j \ge \tau \\
c_i = 0 \Rightarrow \sum_{j \in G_i} s_j < \tau
\end{cases}
\end{equation}
\STATE Select a valid solution $r = (s_1, \dots, s_n)$ from the solution space\;

\STATE \textbf{Step 4: State Vector Application}
\FOR{each anchor point $i$ in $F \cup V$}
    \IF{$s_i = 1$}
        \STATE Apply forward transformation at anchor $i$\;
    \ELSE
        \STATE  Apply reverse transformation at anchor $i$\;
    \ENDIF
\ENDFOR

\RETURN $C'$

\end{algorithmic}
\end{algorithm}

\vspace{-0.3cm}
\subsection{Anchor Points Identification and Grouping}\label{subsec: Anchor Point}

\noindent\textbf{Anchor Points Identification (Alg 1 line 2).} For the given code snippet $C$, we perform static analysis on it to identify anchor points in both channels. Specifically, we extract the set of formal channel anchor points $F$ (based on $T_F$) and the set of natural channel anchor points $V$ (based on $T_V$). We then sort the anchor points in a consistent, fixed order. For $F$, the ordering follows the predefined sequence of rules in $T_F$. For $V$, the ordering is determined by a fixed heuristic. Notably, in the natural channel, some variable names may naturally match one or more renaming variants defined in $T_V$. To ensure consistent ordering during both embedding and extraction, all such variants are first reverted to their untransformed (original) form prior to sorting. We reasonably assume that the sorting function is kept secret from potential attackers.

\noindent\noindent\textbf{Anchor Points Grouping (Alg 1 line 3).} 
For the formal channel, the anchor set $F$ is partitioned into groups of size $\alpha$, i.e., consecutive $\alpha$ anchor points, and the first $n$ groups are selected in order. For the natural channel, all renaming variants of a variable constitute a group, and the first $n$ such groups are likewise selected. Here, $n = \lvert w \rvert$, so that each group corresponds to a bit $w_i$ in the watermark sequence.
A fixed group threshold $\tau \in \mathbb{N}$ is used to define the group state vector $c$. For each group, the group state $c_i$ is directly determined by the corresponding bit $w_i$: $c_i = w_i$.
\vspace{-0.3cm}
\subsection{Adaptive Embedding Capacity Estimation}\label{subsec: Adaptive}
Given the number of available anchor points in \( F \) and \( V \), \tool first estimates the embedding capacity and accordingly selects an appropriate watermark sequence \( w \) and a parity-check matrix \( M \). There are three possible cases:

\noindent\textbf{Case 1: Sufficient anchor points for both grouping and BCH-encoded watermark (Alg 1 line 5-8).}  
If the number of anchor points \( n \) satisfies \( n \geq \alpha \cdot |w| \), it is sufficient for grouping. Here, \( w \) is the BCH-encoded watermark sequence derived from the identity message \( m \).

\noindent\textbf{Case 2: Sufficient anchor points for grouping, but not for the BCH-encoded sequence (Alg 1 line 9-11). }  
If the above condition cannot be satisfied, we fall back to using the identity message directly as the watermark sequence: $w = m$. 
If \( n \geq \alpha \cdot |w| \) is satisfied, grouping can still be performed.

\noindent\textbf{Case 3: Insufficient anchor points for grouping (Alg 1 line 13-16).}  
In extreme cases where the number of anchor points (e.g., 1 or 2) is insufficient even for the minimal grouping requirement, a fixed watermark sequence \( w \) predefined by the organization is used to ensure basic feasibility of detection and verification.
For the first two cases~(Alg 1 line 12), a parity-check matrix \( M \) is selected to match the length of \( w \), satisfying
$\mathrm{rows}(M) = |w|$ and $\quad \mathrm{cols}(M) = |w| \cdot \alpha$. 
For the third case~(Alg 1 line 15.) , a unit (identity) parity-check matrix \( M \) is used.

\vspace{-0.3cm}
\subsection{Watermark Insertion}

Given an LLM-generated code snippet $C$ and an original message $m$, \tool first identifies anchor points as described in Section~\ref{subsec: Anchor Point}, then computes and applies the state vector. 

\noindent\textbf{State Vector Computation (Alg 1 line 18-20).}
In both channels, the problem of determining the state vector is formulated as a Constraint Satisfaction Problem (CSP). Under the grouped state constraints, we solve the following system of linear equations over the binary field:
\begin{equation}
\begin{cases}
M \cdot r \equiv c \pmod{2} \\
c_i = 1 \implies \sum_{j \in G_i} s_j \ge \tau  \\
c_i = 0 \implies \sum_{j \in G_i} s_j < \tau  \\
\end{cases}
\end{equation}

Here, $r$ is the state vector to be determined, $M$ is the parity-check matrix, and $c$ is the grouped state vector derived from the watermark sequence $w$. The solution ensures that the computed anchor point states are consistent with the watermark bit assignments specified at the group level.

Let the solution space of the equation system be denoted by $R$. A solution:
\vspace{-0.2cm}
\begin{equation}
r = (s_1, s_2, \dots, s_n) \in \{0,1\}^n 
\end{equation}
is randomly selected from $R$ and returned as the state vector. If the system has no solution, \tool resamples an alternative watermark sequence $w'$ derived from the same identity message and retries the solving process.

\noindent\textbf{State Vector Application. (Alg 1 line 21-29)}
Given the computed state vectors $r_F$ and $r_V$ for the formal and natural channels respectively, \tool iterates through each anchor point and applies the corresponding transformation. Specifically, for an anchor point with state 1, the associated forward transformation is applied; for a state of 0, the reverse transformation is used to preserve its original form. This process produces the final watermarked code $C'$, which reflects the encoded identity in a semantically equivalent yet syntactically altered version of the original code. 
For example, if $r = (1, 0, 1, 0)$, the first and third anchor points are transformed, while the second and fourth remain unchanged.

\vspace{-0.3cm}
\subsection{Watermark Extraction}

Given a code snippet $C'$ with unknown watermarking status, \tool first identifies anchor points $F$ and $V$ and then preforms the following process.

\noindent\textbf{State Vector Reconstruction.}  
\tool reconstructs the corresponding state vectors \( r_F \) and \( r_V \) using the following rule. For each anchor point in \( F \) or \( V \), its state is determined as:
\begin{equation}
s =
\begin{cases}
1, & \text{if the transformation can be \textit{reversed}}, \\
0, & \text{otherwise}.
\end{cases}
\end{equation}

After evaluating all anchor points, the state vectors \( r_F \) and \( r_V \) are reconstructed as bit sequences, indicating which anchor points are likely to have had transformation rules applied.

\noindent\textbf{Parity-Check Matrix Determination.}  
Based on the number of anchor points, \tool retrieves the parity-check matrices $M_F$ and $M_V$ using the same selection strategy as in Section~\ref{subsec: Anchor Point}. If $C'$ contains an intact watermark, the recovered matrices are expected to match those used during insertion.

\noindent\textbf{Watermark Sequence Derivation.}  
The group state vector $c'$ is derived by applying the fixed thresholding strategy over anchor point groups, following the same grouping scheme in Section~\ref{subsec: Anchor Point}. The recovered watermark sequence $w'$ is computed accordingly.

\noindent\textbf{Watermark Verification.}  
For each channel, \tool checks whether the following condition holds:
\begin{equation}
M \cdot r \equiv c' \pmod{2}
\end{equation}
If the condition is satisfied in at least one channel, $C'$ is deemed to contain a valid watermark. In this case, $w'$ is taken as the embedded watermark sequence, and the corresponding identity message $m'$ can be recovered from $w'$.

%% file: Chapters/evaluation.tex
\vspace{-0.3cm}
\section{Evaluation}

In this section, we present the empirical evaluation of \tool. We begin by introducing the overall experimental setup in Section~\ref{subsec:setup}. Then, we evaluate the detection and extraction accuracy of watermark identification in Section~\ref{subsec:TPR} and Section~\ref{subsec:msgacc}. The robustness of \tool against common code transformations and optimization attacks is analyzed in Section~\ref{subsec:robustness}. Next, we assess the functional transparency of the watermarked code and its impact on usability in Section~\ref{subsec:fidelity}. We further examine the indistinguishability of watermark patterns under statistical analysis in Section~\ref{subsec: indistinguishability}. Finally, Section~\ref{subsec:case study} provides a detailed case study illustrating how \tool behaves under representative real-world attacks.

\vspace{-0.3cm}
\subsection{Setup}
\label{subsec:setup}

\textbf{Baselines.}
We compare \tool with state-of-the-art open source baselines for watermarking LLM-generated code.
\textbf{ACW}~\cite{li2024acw} is a black-box watermarking method. It selectively applies a set of carefully designed semantics-preserving, idempotent code transformations whose presence or absence is used to determine whether a watermark is embedded.
\textbf{SWEET}~\cite{lee2023wrote} is a white-box watermarking approach for code generated by LLMs. It optimizes the decoding procedure of KGW \cite{kirchenbauer2023watermark} by introducing an entropy threshold, thereby encoding watermarks exclusively in high-entropy tokens. 

\noindent\textbf{Datasets.} 
We generate Python code using GPT-4~\cite{openai2023gpt4} across three benchmarks: MBPP~\cite{austin2021program}, APPS~\cite{hendrycksapps2021}, and HumanEval \cite{chen2021codex}. In addition, to enable a comparison with SWEET, we also generate code using StarCoder \cite{li2023starcoder} on the HumanEval and MBPP benchmarks. All datasets contain instruction prompts for code generation, human-written canonical solutions, and test cases for functionality evaluation. Following the approach of Li et al.~\cite{li2024acw}, we remove code snippets with fewer than five lines, as such very short fragments typically have limited value for traceability or copyright analysis. We systematically apply the watermarking scheme to these datasets and compare the results against the original datasets without watermarking.

\textbf{Implementation.}
For the formal channel, we define a set of semantic-preserving transformations, detailed in the replication package~\cite{Replication}. For the natural channel, we define three ordered renaming strategies for each variable, resulting in group sizes of 3 as shown in Table~\ref{tab:variable-conversions}. These independent and reversible transformations allow us to accurately recover anchor states via static analysis during detection, without extra metadata. In contrast, since ACW cannot detect anchors that are naturally in the transformed state, we additionally record the anchors used during embedding and reuse them during watermark detection to improve reliability. For practicality, we treat channels with an all-zero state vector as unwatermarked, since it indicates no transformations were applied. To avoid this case, we never embed messages whose encoded watermark yields an all-zero vector. Instead, we randomly select a nearby valid codeword within the BCH error tolerance to replace it, ensuring the watermark remains detectable.

\begin{table}[ht]
\caption{Available Variable Conversions}
\label{tab:variable-conversions}
\renewcommand{\arraystretch}{1.2}
\begin{tabularx}{\linewidth}{llp{5.5cm}}
\toprule
Name                   & Description                           \\ \midrule
Underline              & Add an underline to the end of the variable.  \\
Initial Capitalization & Capitalize the first letter of a variable. \\
Suffix & \parbox[t]{5.5cm}{Select a suffix from the given list of suffixes based on the context, and add it to the end of the variable.}  \\ 
\bottomrule
\end{tabularx}
\end{table}

We conducted the watermarking operations on StarCoder using a server equipped with eight NVIDIA Tesla V100 SXM2 GPUs (32GB memory each, NVLink). For the GPT-4-related watermarking operations, we used a PC with an 11th Gen Intel(R) Core(TM) i7-11800H CPU (2.30GHz), 16GB DDR4 RAM, running Windows 11.

\vspace{-0.3cm}
\subsection{Watermark Accuracy}
\label{subsec:TPR}

\begin{table*}[]
\caption{Evaluation Results on Discriminability}
\label{tab:discriminability}
\renewcommand{\arraystretch}{1.2}
\begin{tabular}{l|cc|cc|cc|c|c|c}
\hline
            & \multicolumn{2}{c|}{ACW} & \multicolumn{2}{c|}{SWEET} & \multicolumn{2}{c|}{\tool} & \tool-Natural-C & \tool-Formal-C & \tool-Strict \\ \cline{2-10} 
            & TPR(\%)     & FPR(\%)    & TPR(\%)      & FPR(\%)     & TPR(\%)             & FPR(\%)             & TPR(\%)                              & TPR(\%)                             & FPR(\%)                          \\ \hline
APPS-G      & 98.82        & 0.99       & ——           & ——          & 99.64               & 1.34                 & 99.64                                & 99.28                                & 0.42                              \\
HumanEval-G & 99.39       & 0.61       & ——           & ——          & 99.39               & 3.05                 & 99.39                                & 99.39                               & 0.00                              \\
MBPP-G      & 99.79        & 0.00        & ——           & ——          & 100.00               & 0.85                 & 100.00                                & 99.36                                & 0.00                              \\
HumanEval-S & 96.95       & 0.00        & 32.32         & 0.00         & 96.95               & 0.00                 & 96.95                                & 96.95                               & 0.00                              \\
MBPP-S      & 96.40        & 0.00        & 41.46         & 1.00         & 100.00               & 2.20                 & 100.00                                & 98.80                                & 0.00                              \\ \hline
\end{tabular}
\end{table*}

Table~\ref{tab:discriminability} presents the detailed results of \tool and the baselines in attributing the watermark, including TPR and FPR. Specifically, for \tool, we additionally provide the TPR for both the Natural Channel and the Formal Channel, as well as the FPR under strict detection conditions (\tool-Strict, under which the code is considered watermark-free if either channel indicates no watermark). 

Note that the dataset names in the table follow the format of \textit{benchmark-model}. For instance, APPS-G refers to code generated by GPT-4 on the APPS dataset, while MBPP-S denotes code generated by StarCoder on the MBPP dataset.

\noindent\textbf{Overall Results.} \tool demonstrates strong performance across all datasets. For example, on MBPP-G, \tool achieves a TPR of 100.00\%, and on HumanEval-S, a FPR of 0. These results show that \tool can accurately identify the presence of watermarks (in terms of TPR) while effectively avoiding false positives (in terms of FPR). The few cases where watermarks are not correctly detected on Formal Channel are due to edge cases in static analysis, which caused anchor state transformations to fail. 
In our dual-channel watermarking scheme, the FPR is slightly higher due to the backup mechanism, where a watermark is considered present if either channel detects it. While this enhances robustness, it also increases the FPR. To address this, we also evaluate FPR under stricter conditions, where both channels must detect the watermark. Under these conditions, the FPR is much lower, illustrating the trade-off between dual-channel robustness and increased false positives.

\noindent\textbf{Compared to SWEET.} SWEET improves KGW \cite{kirchenbauer2023watermark} by restricting watermark decoding to high-entropy tokens to preserve code quality while maintaining detectability, this approach does mitigate the degradation of code quality caused by altering critical tokens. However, it also reduces the density of ``green'' tokens in the code, thereby weakening the watermark strength. Compared to \tool, the watermark detection TPR decreases by 61.59\% on average.

\noindent\textbf{Compared to ACW.} Compared to ACW, \tool demonstrates consistently high TPR across all datasets, without requiring additional records of the anchors used during watermarking. This improvement can be attributed to the fact that small implementation errors may occur during transformations in practice. ACW determines watermark presence by checking whether applying the same transformations repeatedly yields an identical result on a continuous sequence of several anchors. If error arises during watermark embedding, the watermark becomes undetectable. In contrast, \tool's dual-channel backup mechanism ensures that even if one channel fails, the other still reliably preserve the watermark.

\noindent\textbf{Cross-Organizational Attribution.} Beyond detecting whether a watermark exists, \tool supports cross-organization attribution—the ability to distinguish watermarks applied by different organizations using distinct parity-check matrices. By simply changing the matrix without altering any system-level settings, \tool enables accurate multi-layer tracking of both organizations and individuals. 
To evaluate this, we simulate multiple organizations ($O_1$, $O_2$, $O_3$), each assigned a unique parity-check matrix, and test whether \tool can correctly attribute watermarks to the originating organization ($O$). A higher TPR indicates that these organizations did not mistakenly claim ownership of code produced by $O$.

\begin{table}[]
\centering
\caption{TPRs for Cross-Organization Attribution}
\label{tab:organizations}
\renewcommand{\arraystretch}{1.2}
\begin{tabular}{l|c|c|c}
\hline
            & $O_1$(\%) & $O_2$(\%) & $O_3$(\%) \\ \hline
APPS-G      & 95.89   & 99.67   & 93.35   \\
HumanEval-G & 98.17   & 99.39   & 94.51   \\
MBPP-G      & 96.81   & 100.00  & 95.32   \\
HumanEval-S & 98.17   & 100.00  & 99.39   \\
MBPP-S      & 94.60   & 98.20   & 91.80   \\ \hline
\end{tabular}
\vspace{-0.2cm}
\end{table}

\noindent\textbf{Analysis.} 
Table \ref{tab:organizations} shows that \tool achieves high true positive rates (TPRs) for distinguishing organizational ownership, with values typically above 95\% and reaching 100\% on several datasets, confirming that detectors rarely misidentify watermarks from other organizations as their own. The small variations across organizations (typically 3–5\%) suggest minor overlaps between the encoding spaces of distinct parity-check matrices. Overall, these results demonstrate that \tool effectively preserves cross-organization exclusivity while supporting scalable multi-layer attribution. This validates the scalability of our parity-check–based design for multi-organization deployment without system-level reconfiguration.

\noindent\textbf{Watermark Insertion Overhead.}
We also conduct a preliminary runtime evaluation, and find that watermark embedding and detection incur only modest overhead compared to baselines, since \tool relies solely on lightweight static analysis over the AST, whereas ACW requires invoking a Sourcery API for anchor detection and transformation.

\begin{table}[]
\centering
\caption{Watermark insertion overhead.}
\label{tab:time}
\renewcommand{\arraystretch}{1.2}
\begin{tabular}{c|c|c|c}
\hline
        & \tool & ACW    & SWEET \\ \hline
Time(s) & 0.206                & 16.425 & 0.302 \\ \hline
\end{tabular}
\end{table}

\begin{tcolorbox}

With its dual-channel redundancy, \tool achieves perfect detection (TPR = 100\%). By simply changing the parity-check matrix without altering costly system-level settings, \tool can accurate attribution across both organizational and individual levels.

\end{tcolorbox}

\subsection{Message Accuracy}
\label{subsec:msgacc}
To fairly evaluate the accuracy of watermark sequence recovery, we adopt a simplified BCH code (4,2,1) as used in ACW \cite{li2024acw} for embedding (shown in Table \ref{tab:BCH}), and compare the message accuracy (MsgAcc). MsgAcc is defined as the percentage of watermarks that are correctly recovered after error correction. Since SWEET does not support embedding watermark bit sequences, it is excluded from this evaluation.

For each method, we first identify the set of samples in which embedding a 4-bit message is feasible. Notably, to conserve anchor resources and maximize the applicability of ACW, we omit the dedicated watermark identifier bits and use all available transformation anchors exclusively for embedding the watermark information. MsgAcc is then calculated as the proportion of correctly extracted watermarks over this set of embeddable samples for each scheme, ensuring that the comparison reflects the performance of both methods under equivalent embedding conditions.

Table~\ref{tab:MsgAcc} report the size of the embeddable sample sets (Code-Num) and the detailed results of message recognition for the two methods, respectively.
It is important to note that \tool supports multi-layer attribution, enabling provenance tracing at both the organization and individual levels. When the detector successfully extracts a watermark sequence from the code, the organization-level parity-check matrix identifies which organization applied the watermark, while the $m$ further distinguishes the specific contributor within that organization. This hierarchical design allows \tool to achieve fine-grained ownership tracing without additional system-level reconfiguration. In contrast, ACW is limited to a single-layer scheme that attributes code ownership only at the organizational level and lacks the capability to identify individual contributors.

\begin{table*}[]
\centering
\caption{Example BCH codes used in this work.}
\label{tab:BCH}
\renewcommand{\arraystretch}{1.2}
\begin{tabular}{c|cccc|ccc|ccc|c}
\hline
m            & \multicolumn{4}{c|}{00}                                                                                                   & \multicolumn{3}{c|}{01}                                                                    & \multicolumn{3}{c|}{10}                                                                    & 11                           \\ \hline
BCHEncode(m) & \multicolumn{4}{c|}{0000}                                                                         & \multicolumn{3}{c|}{0101}                                          & \multicolumn{3}{c|}{1010}                                          & 1111 \\ \hline
w            & 0001 & 0010 & 0100 & 1000 & 0101 & 0111 & 1101 & 1010 & 1011 & 1110 & 1111 \\ \hline
\end{tabular}
\end{table*}

\begin{table*}[]
\centering
\caption{Evaluation results on MsgAcc(\%).}
\label{tab:MsgAcc}
\renewcommand{\arraystretch}{1.2}
\begin{tabular}{l|cc|cc|cc|cc|cc}
\hline
 & \multicolumn{2}{c|}{0000} & \multicolumn{2}{c|}{0101} & \multicolumn{2}{c|}{1010} & \multicolumn{2}{c|}{1111}  & \multicolumn{2}{c}{Code-Num}\\ \cline{2-11} 
 & ACW      & \tool      & ACW      & \tool      & ACW      & \tool      & ACW      & \tool     & ACW      & \tool \\ \hline
APPS-G      & 100.00 & 100.00 & 99.46 & 100.00 & 99.73 & 100.00  & 100.00  & 100.00  & 372  & 2249\\ 
HumanEval-G & 100.00  & 100.00 & 97.30 & 100.00 & 100.00  & 100.00  & 100.00  & 100.00  & 37  & 70 \\ 
MBPP-G      & 100.00  & 100.00 & 99.32 & 100.00 & 100.00  & 100.00 & 98.65 & 100.00  & 148 & 295 \\ \hline
\end{tabular}
\end{table*}

\noindent\textbf{Analysis.} The results show that \tool achieves perfect MsgAcc of 100.00\% across almost all datasets and watermark bit sequences, whereas ACW exhibits slightly lower accuracy in a few cases (e.g., 97.30\% on HumanEval-G-0101). This metric reflects the reliability of bit-level watermark extraction; thus, \tool's consistent 100\% MsgAcc demonstrates robust recovery even under diverse code structures.

The minor accuracy drops in ACW are primarily due to potential minor inconsistencies in the implementation of ACW's transformations, which can affect the embedding and recovery of the watermark sequence.  In contrast, \tool's dual-channel encoding provides backup redundancy, allowing successful message reconstruction even if one channel partially fails.

Beyond message accuracy, \tool substantially expands the set of embeddable code instances: supporting approximately six times more samples than ACW on APPS-G and about twice as many on HumanEval-G and MBPP-G. This improvement arises because ACW requires each code snippet to contain at least four available semantic transformations, a condition often unmet in the shorter, cleaner code generated by LLMs. By combining multiple transformation types and variable-renaming anchors, \tool significantly broadens the range of applicable code, enhancing watermark coverage in real-world LLM-generated scenarios.

\noindent\textbf{Long codewords.} 
Following prior watermarking studies~\cite{li2024acw, zhang2025robust, yang2024srcmarker} and considering the limited length of LLM-generated code, our main evaluation focuses on short codewords. To further examine scalability, we extended our evaluation to longer codewords derived from the \textit{(7,4)} BCH scheme, which expands a 4-bit message into a 7-bit codeword while preserving error-correction capability.  We therefore conducted a supplementary experiment using 7-bit codewords on a 760-sample subset of the APPS dataset. The results remained consistent, successfully watermarking 361 files with a TPR of 100\%, an FPR of 0.3\%, and a MsgAcc of 100\%. In contrast, ACW successfully watermarked only three files, as most code samples lacked sufficient anchor points for its embedding mechanism. These results confirm \tool's scalability under longer watermark sequences.

\begin{tcolorbox}
\tool not only maintains perfect message accuracy in watermark recovery but also demonstrates superior applicability across diverse codebases.
\end{tcolorbox}

\subsection{Robustness}
\label{subsec:robustness}
\begin{figure}
    \centering
    \includegraphics[width=\linewidth]{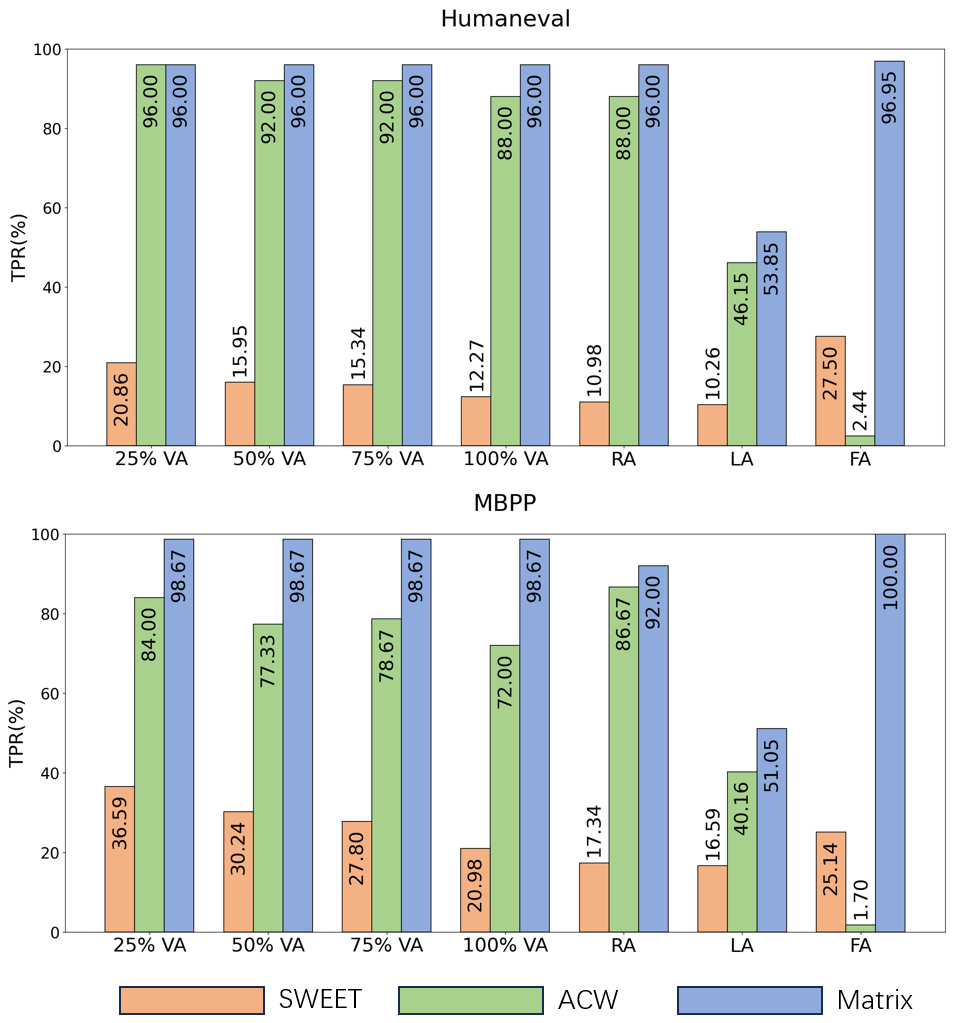}
    \caption{Robustness evaluation results under Variable-rename Attack (VA), Refactor Attack (RA), LLM-based Rewriting Attack (LA), and Reformat Attack (FA).}
    \vspace{-0.5cm}
    \label{fig: detailed atk_results}
\end{figure}

As in Section~\ref{subsec: core-requirements}, we define the attacker as an end user who acquires code from an LLM and seeks to remove embedded watermarks while preserving functionality, with the goal of avoiding attribution and responsibility. We primarily consider four types of attack: (1) Variable Renaming Attack (VA), in which the attacker randomly renames variables in the code; (2) Refactoring Attack (RA), in which the attacker uses a rule-based tool to refactor the watermarked code; (3) LLM-based Rewriting Attack (LA), in which the attacker uses an LLM to rewrite the code; and (4) Reformat Attack (FA), in which the attacker uses formatting tools to sanitize or standardize the code's formatting. We instruct pyrefact~\cite{pyrefact} to refactor the code, GPT-4o~\cite{openai2023chatgpt} to rewrite, and black~\cite{black} to reformat.
Figure~\ref{fig: detailed atk_results} shows a comparison of \tool and the baselines on datasets containing dual-channel watermarks.

\noindent\textbf{Analysis.} As shown in Figure~\ref{fig: detailed atk_results}, \tool consistently outperforms SWEET and ACW across all four attack types on both HumanEval and MBPP datasets.

Under VA, where variable identifiers are randomly replaced, \tool maintains TPRs of 96–98.67\%, far exceeding ACW (72.00\%) and SWEET (20.98\%). This robustness arises from \tool's dual-channel redundancy: when one channel's anchors are perturbed, the other can still recover the watermark. In contrast, ACW's performance drops because one of its watermarking rules depends on sorting variables by name; renaming disrupts that ordering. SWEET is even more vulnerable, as renaming changes the tokenization structure: a variable that originally spanned two tokens might expand to five after renaming, misaligning the token groups used for watermark extraction.

For RA, in which pyrefact restructures the program without altering semantics, \tool again achieves high TPRs (92–98\%), while ACW and SWEET degrade sharply. The dual-channel backup mechanism allows \tool to tolerate such structural transformations. ACW fails because rule-based refactoring invalidates several of its transformation-dependent watermark rules, while SWEET suffers from disrupted token grouping similar to VA.

In the more challenging LA, where GPT-4o regenerates the code with modified syntax, structure, and comments, \tool achieves TPRs of 53.85\% and 51.05\% on HumanEval and MBPP, respectively—about 5–10 percentage points higher than ACW and more than triple SWEET's accuracy. These results reflect the advantage of \tool's parity-check encoding, where the diversity of valid solution spaces enables partial recovery even when both channels are simultaneously perturbed.

Under FA using black, \tool retains nearly perfect TPRs (97–100\%), showing invariance to purely stylistic formatting changes. In contrast, ACW performs poorly because formatting transformations are part of its watermarking rule set; when external tools reformat the code, those transformations produce inconsistent outcomes that invalidate the watermark. SWEET also fails due to token-sequence misalignment, as formatting changes shift the token boundaries used for watermark grouping.

Overall, \tool's robustness derives from two complementary mechanisms: (i) redundant dual-channel encoding, where the natural and formal channels provide mutual backup, and (ii) solution-space diversity in the parity-check system, ensuring that perturbed state vectors that remain within the valid subspace still yield correct watermark recovery.

\begin{tcolorbox}
Benefiting from the diverse solution space of its parity-check encoding and the redundancy of its dual-channel design, \tool demonstrates strong robustness against Variable-Renaming and Rule-based Refactoring Attacks, and maintains reliable detection under the more challenging LLM-based Rewriting Attack. Moreover, its format-independent watermarking mechanism ensures complete immunity to Reformat Attacks.
\end{tcolorbox}

\subsection{Fidelity}
\label{subsec:fidelity}
To verify that watermarking does not compromise code functionality, we conduct an experiment comparing the unit test pass rates of code before and after watermark embedding. We partition the dataset into unwatermarked and watermarked code samples. It is important to note that, due to the limited generation capabilities of LLMs, the pass rate of the original generated code was not 100.00\%.

\noindent\textbf{Analysis.} As shown in Table~\ref{tab:passrate}, \tool effectively preserves the functionality of the generated code, achieving negligible degradation across all datasets. For example, on APPS, the pass rate drops by only 0.14\%, and remains completely unchanged on HumanEval and MBPP, confirming that the embedding and extraction procedures introduce no semantic or syntactic disruption. This demonstrates that \tool maintains code correctness even under extensive watermarking.

In comparison, ACW shows slightly higher but still acceptable degradation (under 0.2\%), since small implementation errors may occur during transformations in practice. SWEET, however, experiences noticeable degradation: up to 7.60\% on MBPP-S and 2.44\% on HumanEval-S. The performance drop originates from its watermarking mechanism, which injects constraints into the LLM decoding process. Although SWEET avoids watermarking low-entropy or critical tokens, the lack of a global syntactic view can cause inconsistencies between modified and unmodified code regions, leading to broken dependencies and failed test cases.

\begin{tcolorbox}
\tool uses semantically equivalent code transformations to embed the watermark, ensuring near-lossless embedding with minimal impact on code functionality.
\end{tcolorbox}

\begin{table}[]
\caption{Evaluation results on Pass Rate.}
\label{tab:passrate}
\renewcommand{\arraystretch}{1.2}
\begin{tabular}{l|l|cc|c}
\hline
\multirow{2}{*}{}                     & \multirow{2}{*}{Dataset} & \multicolumn{2}{c|}{Pass   Rate(\%)}         & \multicolumn{1}{l}{\multirow{2}{*}{Degradation(\%)}} \\ \cline{3-4}
                                      &                          & \multicolumn{1}{c|}{Before} & After & \multicolumn{1}{l}{}                                 \\ \hline
\multirow{5}{*}{ACW}                  & APPS-G                   & \multicolumn{1}{c|}{68.01}     & 67.81       & 0.20                                                 \\
                                      & HumanEval-G              & \multicolumn{1}{c|}{82.32}     & 82.32       & 0.00                                                    \\
                                      & MBPP-G                   & \multicolumn{1}{c|}{71.28}     & 71.28       & 0.00                                                    \\
                                      & HumanEval-S              & \multicolumn{1}{c|}{35.98}     & 35.98       & 0.00                                                    \\
                                      & MBPP-S                   & \multicolumn{1}{c|}{50.40}     & 50.40       & 0.00                                                    \\ \hline
\multirow{2}{*}{SWEET}                & HumanEval-S              & \multicolumn{1}{c|}{35.98}     & 33.54       & 2.44                                                 \\
                                      & MBPP-S                   & \multicolumn{1}{c|}{50.40}     & 42.80       & 7.60                                                 \\ \hline
\multirow{5}{*}{\tool} & APPS-G                   & \multicolumn{1}{c|}{68.01}     & 67.87       & 0.14                                                 \\
                                      & HumanEval-G              & \multicolumn{1}{c|}{82.32}     & 82.32       & 0.00                                                    \\
                                      & MBPP-G                   & \multicolumn{1}{c|}{71.28}     & 71.28       & 0.00                                                    \\
                                      & HumanEval-S              & \multicolumn{1}{c|}{35.98}     & 35.98       & 0.00                                                    \\
                                      & MBPP-S                   & \multicolumn{1}{c|}{50.40}     & 50.40       & 0.00                                                    \\ \hline
\end{tabular}
\end{table}

\vspace{-0.2cm}
\subsection{Indistinguishability under Statistical Analysis}
\label{subsec: indistinguishability}

As discussed in Section~\ref{subsec: core-requirements}, a strong adversary may obtain multiple watermarked variants of the same code and analyze anchor activation patterns to infer anchor–bit mappings, enabling possible watermark removal or manipulation. We therefore evaluate \tool's resistance to such statistical inference by comparing its activation patterns and intra-watermark diversity against ACW.

We define $\epsilon$-indistinguishability: For any two watermark messages $m_0$ and $m_1$, the distributions of anchor activations $P(A\mid m_0)$ and $P(A\mid m_1)$ are indistinguishable if, for any adversary $A$,
$\left| \Pr[\text{Attack}_A(m_0)] - \Pr[\text{Attack}_A(m_1)] \right| \leq \epsilon$, 
where $\epsilon$ represents the maximum distinguishing advantage. Smaller $\epsilon$ indicates stronger indistinguishability.

\begin{figure}
    \centering
    \includegraphics[width=\linewidth]{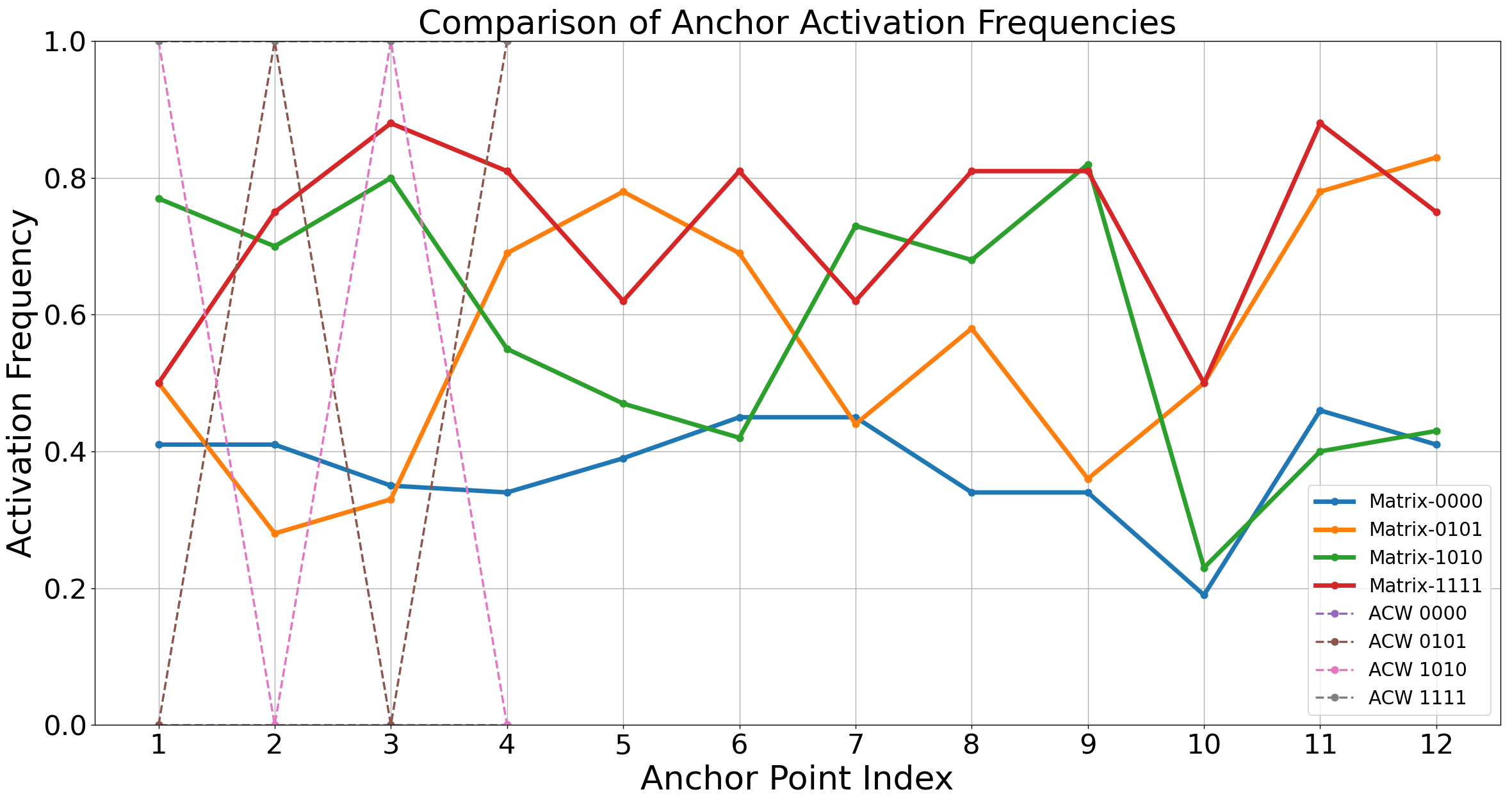}
    \caption{Activation Frequency of \tool, Each curve represents a different watermark, with the y-axis showing the frequency of anchor activation (state 1) across all samples.}\vspace{-0.5cm}
    \label{fig: Activation Frequency}
\end{figure}

\noindent\textbf{Anchor activation analysis.} Figure~\ref{fig: Activation Frequency} reports the activation frequencies of 12 anchors under four watermark messages (0000, 0101, 1010, 1111). For ACW, each bit deterministically drives a designated anchor (e.g., ``1010'' triggers anchors 1–4). This one-to-one mapping yields near-binary activation curves that are clearly separable across messages, exposing highly predictable statistical structure that an adversary could readily exploit to recover the bit–anchor mapping.

By contrast, \tool exhibits stochastic and overlapping activation behavior. Anchor activation frequencies vary smoothly between roughly 0.3 and 0.9 across watermark messages, and no anchor consistently responds to any particular bit. To quantify $\epsilon$-indistinguishability, we compare the activation distributions using Jensen–Shannon divergence (JSD) and mutual information (MI). Across 12 anchors and 6 message pairs (72 comparisons), the average JSD is only 0.0479 (max 0.1704), indicating substantial overlap between distributions. The MI is similarly low (avg. 0.0479 nats), implying that all anchors combined leak $<$ 1 bit about the 4-bit watermark—insufficient for any meaningful message inference. These results show that \tool substantially limits statistical distinguishability: anchor activations reveal no stable bit-dependent patterns, and the information available to an adversary is far below the threshold needed to reconstruct or even partially infer the embedded watermark.

\begin{figure}
    \centering
    \includegraphics[width=\linewidth]{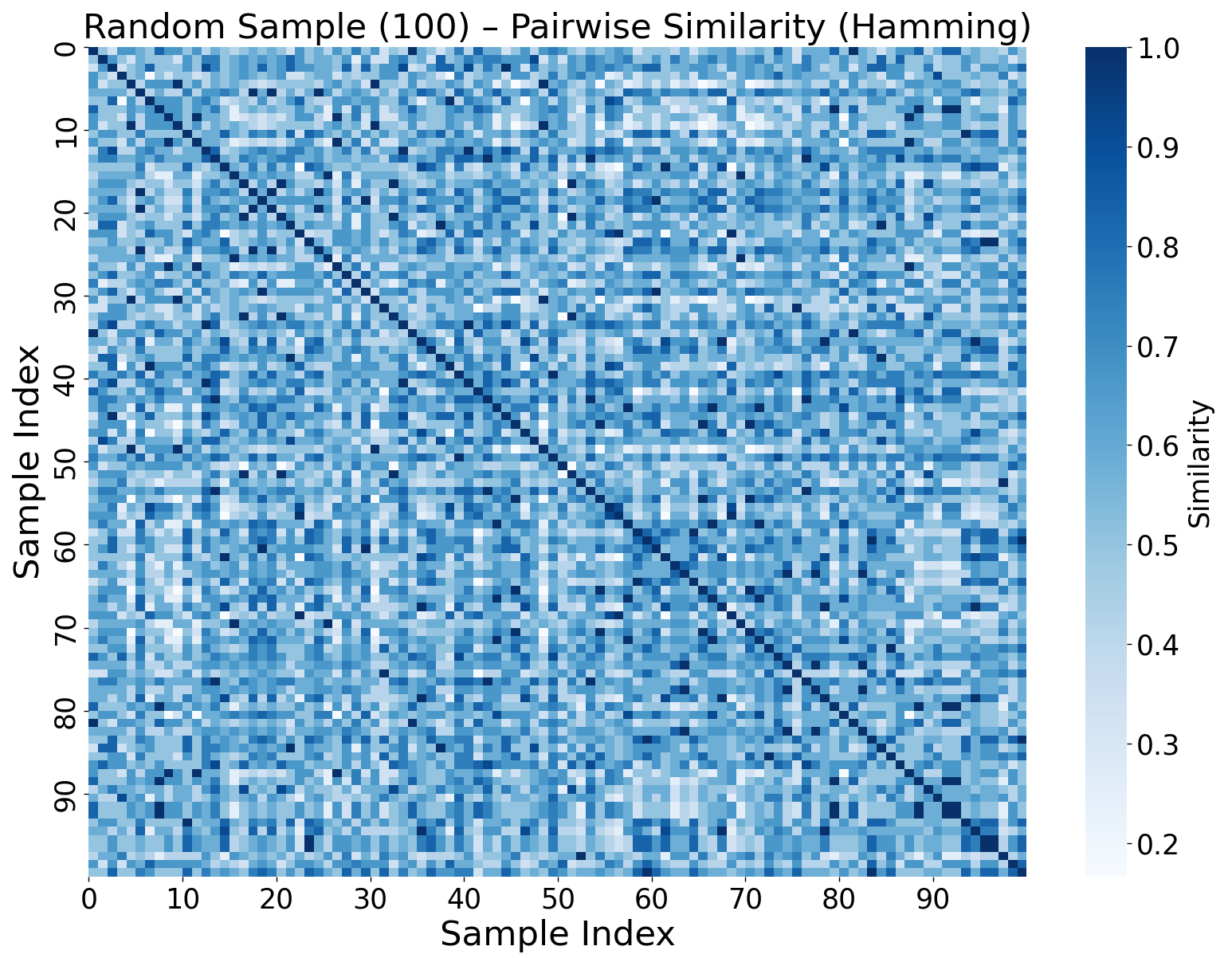}
    \caption{Pairwise similarity heatmap of \tool}
    \label{fig: Similarity heatmaps}
    \vspace{-0.5cm}
\end{figure}

\noindent\textbf{Intra-watermark diversity.} Figure \ref{fig: Similarity heatmaps} evaluates whether an adversary, after collecting many samples carrying the same watermark, could identify stable structural signatures. The heatmap plots the pairwise Hamming similarity of 100 watermarked samples. Instead of collapsing to a single canonical pattern, the similarities span a broad range (0.3–0.8), indicating that \tool produces many distinct state vectors for the same message.

This variability stems from the randomized BCH mapping and the multiplicity of valid solutions to the parity-check constraints. Crucially, this means an adversary cannot isolate invariant features or derive a reliable template characterizing a watermark message—even when the watermark bits are fixed. Unlike ACW, which yields one deterministic state vector per message, \tool provides no consistent structural anchor that could support clustering or pattern-matching attacks.

These results show that \tool preserves strong intra-message unpredictability: repeated samples leak no stable bit-dependent structure and therefore offer no actionable signal for reconstructing, distinguishing, or reverse-engineering watermark assignments. This statistical dispersion acts as an additional defense layer beyond activation-distribution indistinguishability, further limiting the adversary's ability to infer internal encoding rules from observation alone.

\begin{tcolorbox}
\tool's checksum-matrix construction and randomized BCH mapping reduce anchor-activation predictability, improving resistance to statistical inference. Even with many samples, adversaries cannot reliably distinguish watermark-induced transformations from natural code variations.
\end{tcolorbox}

\vspace{-0.4cm}
\subsection{Case Study}
\label{subsec:case study}
While the previous sections evaluate \tool statistically, here we present a detailed case study to illustrate how it behaves under concrete attack instances.

\begin{figure*}[h]
    \centering
    \subfloat[Attack to Formal channel]{
        \includegraphics[width=0.9\textwidth]{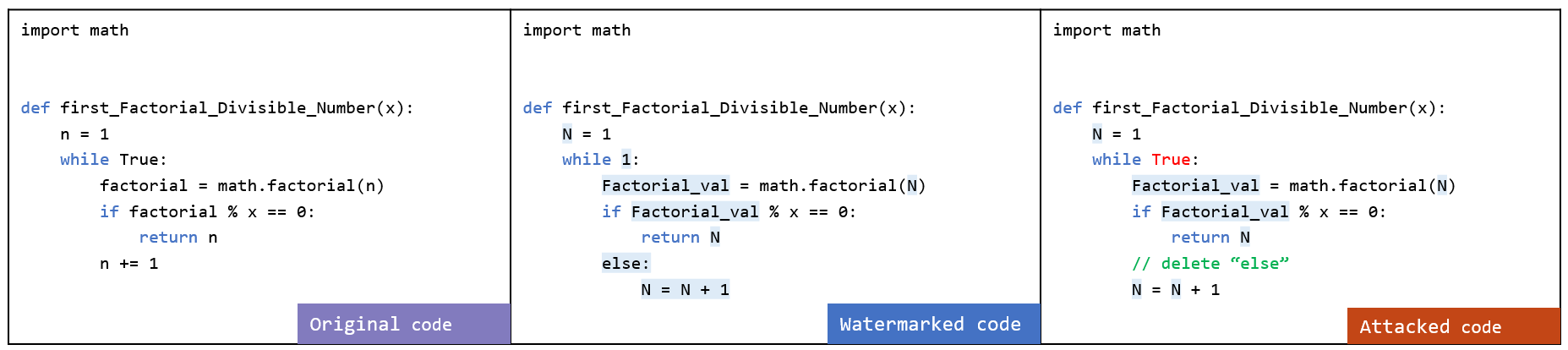}
        \label{fig: success case Formal channel}
    }
    \vspace{-0.2cm}

    \subfloat[Attack to Natural channel]{
        \includegraphics[width=0.9\textwidth]{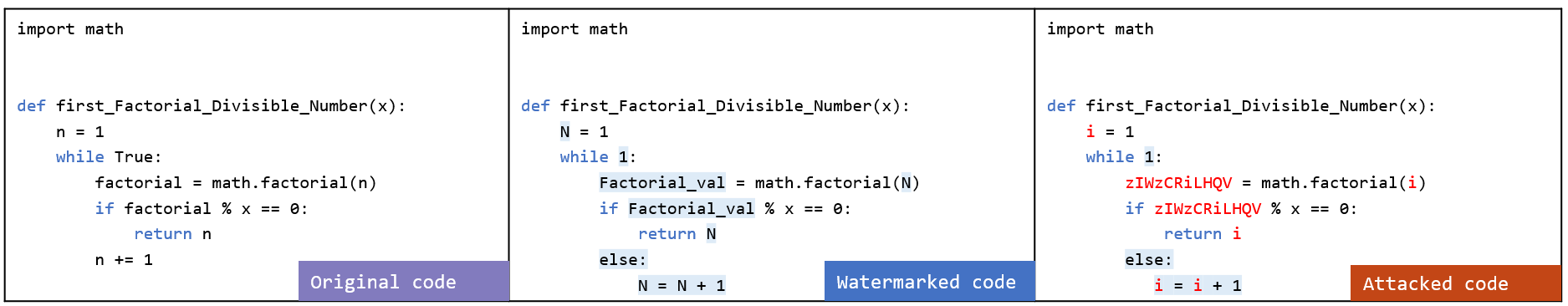}
        \label{fig: success case Natural channel}
    }
    \vspace{-0.2cm}
    \subfloat[Attack to Both channel]{
        \includegraphics[width=0.9\textwidth]{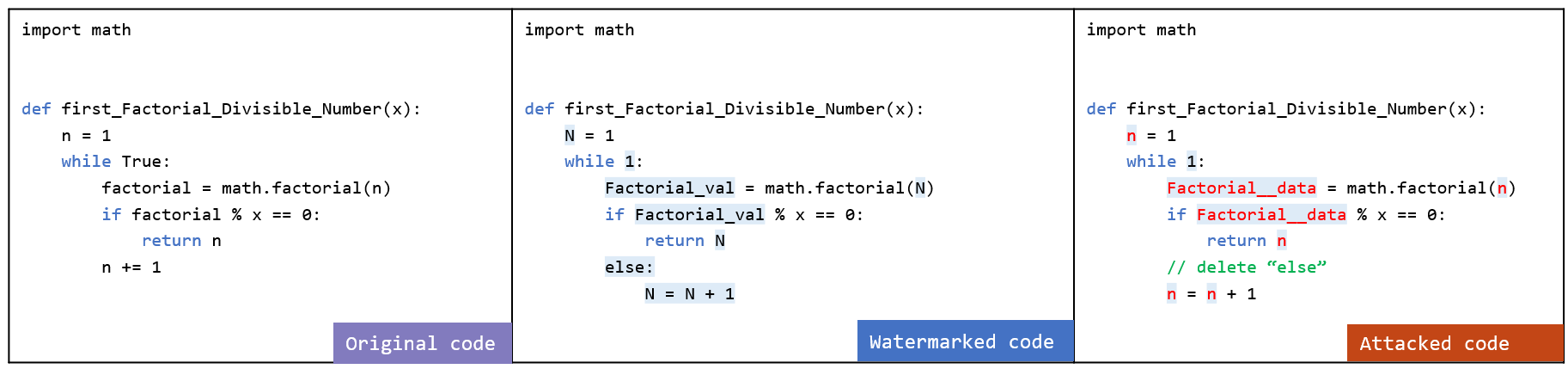}
        \label{fig: success case Both channel}
    }
    
    \caption{Case Study: The red parts are the attacked code, and the blue shaded areas are the watermarked code.}
    \label{fig: cases}
    \vspace{-0.5cm}
\end{figure*}

Figures~\ref{fig: success case Formal channel} and~\ref{fig: success case Natural channel} illustrate \tool's robustness under single-channel attacks. In these cases, the adversary heavily rewrites one channel: either by renaming most variables in the formal channel or by altering sentence structures in the natural channel. However, the other channel remains unaffected and still provides a valid anchor state vector that satisfies the parity-check constraints. Since \tool embeds codeword redundantly across both channels, the decoder can recover the watermark from the intact channel, demonstrating that severe corruption of one channel does not compromise watermark extraction.

Figure~\ref{fig: success case Both channel} illustrates the resilience of \tool under simultaneous dual-channel attacks. In this scenario, the adversary rewrites nearly all variable names in the formal channel and completely disrupts the natural-channel structures. Despite these extensive modifications, the resulting anchor state vector remains within the feasible solution space defined by the verification matrix. This indicates that the parity-check constraints provide sufficient redundancy to absorb substantial structural and lexical perturbations, allowing the decoder to recover the original watermark. Thus, \tool succeeds even under compound attacks that target both embedding channels.

A detailed analysis is provided in our replication package.

%% file: Chapters/discussion.tex
\vspace{-0.3cm}
\section{Discussion}
\subsection{Practicality Analysis}

\tool is theoretically sound, but practical deployment may face several challenges. Detection failures can occur in rare cases where static analysis is hindered by complex or unconventional code, though the dual-channel backup helps mitigate this risk. Attribution ambiguity may also arise if watermark patterns are insufficiently unique, potentially leading to mistaken ownership claims.

Unit tests verify functional correctness, but cannot guarantee full semantic preservation under all edge cases. Moreover, a highly informed adversary might still craft strategies to circumvent detection, despite the system’s randomized transformations and parity-check encoding.

Nevertheless, these limitations are minor in typical use cases, and \tool remains robust and practical for real-world adoption.

\vspace{-0.3cm}
\subsection{Limitations}
While \tool provides a robust and flexible solution for code provenance, several aspects could benefit from further exploration. 

\noindent\textbf{Granularity requirement for detection.} Like prior schemes~\cite{li2024acw, zhang2025robust, yang2024srcmarker}, \tool requires syntactically valid code for detection, as static analysis cannot be performed on ill-formed or incomplete code fragments. While practical instances show that \tool can be extended to embed watermarks at the file level, detecting watermarks in arbitrary fragments, especially those lacking syntactic integrity, remains challenging and would require fundamentally different techniques. Nevertheless, function-level or block-level code is sufficient for most practical scenarios where complete units of code, rather than isolated lines, are reused.

\noindent\textbf{Scalability of parity-based encoding scheme.} The number of independently verifiable organizations is bounded by available transformation anchors. While our dual-channel design expands capacity 2-6× over baselines, extremely fine-grained multi-organization scenarios may exceed anchor limits. Nevertheless, our current capacity suffices for most practical deployments with reasonable numbers of stakeholders.

\noindent\textbf{Reliance on manually curated set of code transformations.}
Our semantic-preserving transformations use handcrafted rules that ensure correctness but may not cover all coding styles. While learning-based approaches could expand coverage, they sacrifice interpretability and robustness to distribution shifts. Our rule-based design prioritizes reliability and transparency, which is critical for security applications, over exhaustive style coverage.

\vspace{-0.4cm}

%% file: Chapters/literature.tex
\vspace{-0.4cm}
\section{Related Work}\label{sec: Related work}
\subsection{Software Watermarking}
Software watermarking is a research area dedicated to embedding secret signals into code to protect the owner's copyright without degrading performance. This field is generally categorized into dynamic and static watermarking approaches.

Dynamic watermarking methods \cite{wang2018exception, ma2019xmark} inject watermarks during compilation or execution. For example, by triggering hidden control flows or specific execution states during operation \cite{chen2017hidden, tian2015software, wang2018exception}, enabling the watermark to be extracted and verified at runtime.

Among static watermarking techniques, one approach embeds watermarks as part of the software by transforming executables \cite{balachandran2014function, chen2018software, cho2014learning}. For instance, Kang et al. \cite{kang2021softmark} encoded watermarks by reordering binary functions, while Monden et al. \cite{monden2000practical} inserted virtual methods carrying watermark bitstrings into Java class files.

However, for copyright protection of LLM-generated code, source code is usually released as plain text visible to developers. Thus, static methods based on reordering, virtual methods, or dead code may be conspicuous, making direct watermarking of source code text a more suitable strategy, similar to natural language watermarking.

\vspace{-0.3cm}
\subsection{Black-box Code Watermarking}

Tosyn~\cite{li2023protecting} resists imitation attacks by shifting code syntax away from the model's training distribution through rule-based transformations. ACW~\cite{li2024acw} uses idempotent, semantics-preserving transformations, but its reliance on all \(n \ge 6\) rules limits robustness, while fixed identifier bits constrain capacity and complicate recovery. More generally, these transformation-based methods often produce isomorphic syntactic patterns, making watermarks vulnerable to differential detection, theft, and forgery.

CodeMark~\cite{li2023towards} embeds bit strings into variable names, preserving both natural and operational semantics using a contextual scheme with graph neural networks.
SrcMarker~\cite{yang2024srcmarker} uses a dual-channel Transformer to embed watermarks in code structure and variable names. While preserving functionality, its reliance on a shallow model limits detection performance due to weak feature extraction. 
RoSeMary~\cite{zhang2025robust} integrates code watermarking with zero-knowledge proofs to verify code provenance without revealing the watermark.
These learning-based watermarking often face challenges in  explainability, and deployment practicality. 

\vspace{-0.5cm}
\subsection{White-box Code Watermarking}

SWEET~\cite{lee2023wrote} partitions high-entropy tokens into green and red lists to restrict LLM sampling to “safe” tokens. CodeIP~\cite{guan2024codeip} predicts the syntactic roles of tokens to determine suitable embedding locations. STONE~\cite{kim2025marking} identifies non-syntactic tokens via entropy and introduces evaluation metrics such as detectability and naturalness. STA-M~\cite{mao2024watermark} employs soft entropy-based thresholds to steer token sampling without hard constraints. MCGMark~\cite{ning2024mcgmark} embeds watermarks with detection and error-correction bits, using syntax- and structure-aware token skipping for robustness.
These approaches rely on local heuristics rather than a global understanding of code syntax and semantics, making them prone to subtle errors or functionality degradation. Moreover, most support only watermark detection, rather than arbitrary message embedding, which limits their use in advanced attribution and tracking scenarios.

%% file: Chapters/conclusion.tex
\vspace{-0.6cm}
\section{Conclusion}

In this work, we present \tool, a secure watermarking framework that balances detectability, fidelity, robustness, and interpretability. Our method combines grouping constraints, BCH error-correcting codes, and parity-check matrices to improve robustness and detectability while keeping watermark fingerprints statistically concealed. We further employ semantic-preserving code transformations and variable renaming to maintain functional correctness, and use a dual-channel mechanism for mutual backup to further enhance robustness. Extensive experiments on diverse code benchmarks demonstrate the effectiveness of \tool.

\vspace{-0.2cm}